\documentstyle[12pt]{article}
\begin{document}
\tolerance=5000
\def\be{\begin{equation}}
\def\ee{\end{equation}}
\def\bea{\begin{eqnarray}}
\def\eea{\end{eqnarray}}
\def\nn{\nonumber \\}
\def\cF{{\cal F}}
\def\det{{\rm det\,}}
\def\Tr{{\rm Tr\,}}
\def\e{{\rm e}}
\def\etal{{\it et al.}}
\def\erp2{{\rm e}^{2\rho}}
\def\erm2{{\rm e}^{-2\rho}}
\def\er4{{\rm e}^{4\rho}}
\def\etal{{\it et al.}}

\ 

\vskip -2cm

\ \hfill
\begin{minipage}{3.5cm}
NDA-FP-57 \\
March 1998 \\
\end{minipage}

\vfill

\begin{center}
{\Large\bf On the conformal anomaly from higher derivative gravity 
in AdS/CFT correspondence}

\vfill

{\sc Shin'ichi NOJIRI}\footnote{\scriptsize 
e-mail: nojiri@cc.nda.ac.jp, snojiri@yukawa.kyoto-u.ac.jp} and
{\sc Sergei D. ODINTSOV$^{\spadesuit}$}\footnote{\scriptsize 
e-mail: odintsov@mail.tomsknet.ru, odintsov@itp.uni-leipzig.de}

\vfill

{\sl Department of Mathematics and Physics \\
National Defence Academy, 
Hashirimizu Yokosuka 239, JAPAN}

\ 

{\sl $\spadesuit$ 
Tomsk Pedagogical University, 634041 Tomsk, RUSSIA \\
}

\ 

\vfill

{\bf abstract}

\end{center}

We follow to Witten proposal \cite{W} in the calculation 
of conformal anomaly from $d+1$-dimensional higher derivative 
gravity via AdS/CFT correspondence. It is assumed 
that some $d$-dimensional conformal field theories have a description
in terms of above $d+1$-dimensional higher derivative gravity 
which includes not only Einstein term and cosmological constant 
but also curvature squared terms. The explicit expression 
for two-dimensional and four-dimensional anomalies is found,
it contains higher derivative corrections. In particular,
it is shown that not only Einstein gravity but also
theory with the Lagrangian $L=aR^2+bR_{\mu\nu}R^{\mu\nu} 
+ \Lambda$ (even when $a=0$ or $b=0$) is five-dimensional 
bulk theory for $d=4$ ${\cal N}=4$ super Yang-Mills theory 
in AdS/CFT correspondence. Similar $d+1=3$ theory with
(or without) Einstein term may describe $d=2$ scalar or
spinor CFTs. 
That gives new versions of bulk side 
 which may be useful in different aspects. As application 
of our general formalism we find next-to-leading corrections 
to the conformal anomaly of ${\cal N}=2$ supersymmetric theory 
from $d=5$ AdS higher derivative gravity (low energy string
 effective action).

\section{Introduction. Conformal anomaly from dilatonic gravity.}

Recently so-called AdS/CFT (anti-de Sitter space vs. 
conformal field theory) correspondence attracted a lot of  
attention \cite{Mal,GKP,W}. In particulary, 
the correspondence between correlation functions in AdS and 
those of the boundary manifold was discussed.

$D$ dimensional anti-de Sitter space can be realized by 
imposing a constraints on $D+1$ coordinates:
\be
\label{II1}
 - x_0^2 + x_1^2 + \cdots x_{D-1}^2 - x_D^2= -L^2 \ .
\ee
 From this realization, it is easy to see that this space 
has $SO(D-1,2)$ symmetry as isometry. 
The algebra of $SO(D-1,2)$ symmetry is  the same as the algebra 
of conformal transformations acting on $D-1$ dimensional 
Minkowski space. 

In an adequate coordinate choice, the metric on the 
$D$-dimensional anti-de Sitter space is given by
\be
\label{II3}
ds_{\rm AdS}^2=\rho^{-2}d\rho^2 + \rho^{-1}\sum_{i,j=0}^{D-2}\eta_{ij}
dx^idx^j\ .
\ee
Here $\eta$ is the metric on the flat $D-1$-dimensional Minkowski 
space. We should note that there is a boundary when $\rho$ 
vanishes. The topology of the boundary is almost that of the 
$D-1$ dimensional Minkowski space, or more exactly, 
Minkowski space with a point at infinity, that is, 
topologically compactified Minkowski space.
On the boundary manifold, $SO(D-1,2)$ acts exactly as usual 
conformal transformation. When we consider the surface with 
fixed finite $\rho$, there is a correction proportional $\rho$ 
only in the conformal boost but the correction vanishes just on 
the boundary, that is, in the limit that $\rho$ vanishes.

AdS/CFT correspondence is conjectured in ref.\cite{Mal}. 
When $N$ p-branes in superstring theory or so-called M-theory 
coincide with each other and the coupling constant is small, 
the classical supergravity on AdS${_{D=d+1=p+2}}$, which is 
the low energy effective theory of superstring or M-theory, 
is, in some sense, dual to large $N$ conformal field theory 
on $M^d$, which is the boundary of the AdS.
For example, $d=2$ case corresponds to (4,4) 
superconformal field theory, 
$d=4$ case corresponds to $U(N)$ or $SU(N)$ ${\cal N}=4$ 
super Yang Mills theory and $d=6$ case to (0,2) 
superconformal field theory. 

The conjecture tells that partition function in 
$d$-dimensional conformal field theory is given in terms 
of the classical action in $d+1$-dimensional gravity theory:
\be
\label{II4}
Z_d(\phi_0)=\e^{-S_{\rm AdS}\left(\phi^{\rm classical}
(\phi_0)\right)}\ .
\ee
Here $\phi_0$ is the value of the field $\phi$ on the boundary and 
$\phi^{\rm classical}(\phi_0)$ is a field on bulk background, 
which is AdS, given by solving the equations of motion with 
the boundary value $\phi_0$ on $M^d$.
$S_{\rm AdS}\left(\phi^{\rm classical}(\phi_0)\right)$ 
is the classical gravity action on AdS. 
When we substitute the classical solution into the action, 
the action, in general, contains infrared divergences 
coming from the infinite volume of AdS.
Then we need to regularize the infrared divergence. 
It is known that as  
 a result of the regularization and the renormalization
(for the introduction to the renormalization in background field 
method see ref.\cite{BOS}) 
there often appear anomalies.  In ref.\cite{W}
Witten made the proposal how to calculate the conformal 
anomaly from classical gravity (bulk) side. This proposal has been 
worked out in detail in ref.\cite{HS}
(for Einstein gravity) where 
 it was shown 
that the conformal (Weyl) anomaly may be recovered after 
 regularizing the 
above infrared divergence (from bulk side-
Einstein theory). The usual result  for 
conformal anomaly of boundary QFT thus may be reproduced.
This is a kind of IR-UV duality.

Since the calculation of ref. \cite{HS} was done 
without dilaton background, in the previous paper \cite{NOads}, 
the authors investigate the 
conformal anomaly in the non-trivial dilaton background.
We start from 
the action of $d+1$-dimensional dilatonic gravity with boundary 
terms
\bea
\label{II5}
S&=&{1 \over 16\pi G}\left[\int_{M_{d+1}} d^{d+1}x \sqrt{-\hat G}
\left\{ \hat R + X(\phi)(\hat\nabla\phi)^2 + Y(\phi)\hat\Delta\phi
+ 4\lambda^2  \right\}\right. \nn
&& \left. +\int_{M_d} d^dx\sqrt{-\hat g}
\left(2\hat\nabla_\mu n^\mu + \alpha\right)  \right]\ .
\eea
Here $M_{d+1}$ is $d+1$ dimensional manifold, which is 
identified with AdS${_{d+1}}$ and $n_\mu$ is the unit 
vector normal to the boundary manifold $M_d$. 
$\alpha$ is a parameter which is chosen properly. 
 The boundary terms play a role for cancellation of  
the leading infrared divergences which guarantee 
the system depends only on the boundary value. Note 
that dilatonic gravity makes the beatiful realization 
of two-boundaries AdS/CFT correspondence \cite{NOA} 
(for other examples of such two-boundaries correspondence 
see refs.\cite{BST,Be}).   

$X(\phi)$, $Y(\phi)$ can be arbitrary functions of $\phi$. 
 The arbitrariness is not real but apparent. In fact, by 
the redefinition, 
\be
\label{II6}
 \varphi\equiv\int d\phi\sqrt{2V(\phi)}\ ,\ \ 
V(\phi)\equiv X(\phi)-Y'(\phi) 
\ee
the action can be rewritten as
\bea
\label{II7}
S&=&{1 \over 16\pi G}\left[\int d^{d+1}x \sqrt{-\hat G}
\left\{ \hat R + 2(\hat\nabla\varphi)^2 + 4\lambda^2  
\right\}\right. \nn
&& \left. +\int_{M_d} d^dx\sqrt{-\hat g}
\left(2\hat\nabla_\mu n^\mu 
+ \alpha + Y(\phi)n^\mu\partial_\mu\phi \right) \right]\ .
\eea
The last term on $M_d$ finally does not contribute to anomaly.

 Following the calculation in \cite{HS}, the explicit expression 
for the conformal anomaly has been obtained 
\be
\label{II8}
T={l \over 16\pi G} \left\{ R_{(0)}
+ X(\phi_{(0)})(\nabla\phi_{(0)})^2 
+ Y(\phi_{(0)})\Delta\phi_{(0)}
\right\}
\ee
for $d=2$. This result could be compared with the UV-calculation
of the conformal anomaly of matter 
dilaton coupled $N$ scalars \cite{E,NO,BH} 
and $M$ Majorana dilaton coupled  spinors \cite{NNO}
 whose action is given by 
\be
\label{II9}
S={1 \over 2}\int d^2x \sqrt{-g}\left\{
f(\phi) g^{\mu\nu}\sum_{i=1}^N\partial_\mu\chi_i
\partial_\nu\chi_i + g(\phi) \sum_{i=1}^M
\bar\psi_i\gamma^\mu\partial_\mu\psi_i\right\}\ .
\ee
The results in (\ref{II8}) correspond to
\bea
\label{II10}
&& {l \over 16\pi G} ={2N + M \over 48\pi} \ ,\nn 
&& {l \over 16\pi G}X(\phi_{(0)})=-{N \over 4\pi}
\left({f'' \over 2f}- {{f'}^2 \over 4f^2}\right) 
-{M \over 24\pi}\left({g'' \over g} - {{g'}^2 \over g^2}\right)
\ ,\nn
&& {l \over 16\pi G}Y(\phi_{(0)})=-{N \over 4\pi}{f' \over 2f}
- {M \over 24\pi}{g' \over g}\ .
\eea
For $d=4$, the expression of the conformal anomaly 
from bulk dilatonic gravity has the following 
form:
\bea
\label{II11}
T&=&{l^3 \over 8\pi G} 
\left[ {1 \over 8}R_{(0)ij}R_{(0)}^{ij}
-{1 \over 24}R_{(0)}^2 \right. \nn
&& + {1 \over 2} R_{(0)}^{ij}\partial_i\varphi_{(0)}
\partial_j\varphi_{(0)} - {1 \over 6} R_{(0)}g_{(0)}^{ij}
\partial_i\varphi_{(0)}\partial_j\varphi_{(0)}  \nn
&& \left. + {1 \over 4}
\left\{{1 \over \sqrt{-g_{(0)}}} \partial_i\left(\sqrt{-g_{(0)}}
g_{(0)}^{ij}\partial_j\varphi_{(0)} \right)\right\}^2 + {1 \over 3}
\left(g_{(0)}^{ij}\partial_i\varphi_{(0)}\partial_j\varphi_{(0)} 
\right)^2 \right]\ .
\eea
In \cite{LT}, it has been calculated the 
conformal anomaly coming from the multiplets of ${\cal N}=4$ 
supersymmetric $U(N)$ or $SU(N)$ Yang-Mills coupled with 
${\cal N}=4$ conformal supergravity(in a covariant way).
If we choose the length parameter $l$ as
\be
\label{II13}
{l^3 \over 16\pi G}={2N^2 \over (4\pi)^2}
\ee
and consider the background where only gravity and the real part 
of the scalar field $\varphi$ are non-trivial and other 
fields vanish, IR-calculation (\ref{II11}) from the bulk
 exactly reproduces the 
 result of UV-calculation by Liu and Tseytlin.
Since dilaton field always appears in string theory, 
the conformal anomaly gives further check of 
AdS/CFT correspondence.

In this paper, we consider $R^2$-gravity, which contains the squares 
of the curvatures in the action, as an extension of Einstein gravity.
This is our bulk theory (for a general review and list of refs.
 on quantum higher derivative gravity see\cite{BOS}). Moreover 
for some combinations of higher derivative terms it corresponds 
to different compactifications of superstring and heterotic string 
effective actions.

The paper is organised as follows. In the next section we present 
general formalism
for the calculation  of conformal anomaly from bulk higher derivative 
gravity. As an example we give the explicit result for two-dimensional
CFT anomaly  
as obtained from three-dimensional higher derivative gravity.
Section three is devoted to the same calculation in four-dimensional case.
We find  conformal $d=4$ 
 QFT (boundary ${\cal N}=4$ super Yang-Mills theory)
 which has description 
in terms of five-dimensional higher derivative gravity. 
The application of results of section 3 
to evaluation of next-to-leading correction 
to ${\cal N}=2$ supersymmetric theory conformal 
anomaly from bulk AdS higher derivative gravity 
(low energy string effective action) is done in
 section 4. The last
section is devoted to brief description of some perspectives. 

\section{Conformal anomaly from bulk higher derivative gravity.
General formalism and d2 case.}

As an extension of the Einstein gravity, we consider $R^2$ 
gravity (bulk theory),
 which contains the squares of the curvatures in the action:
\bea
\label{vi}
S&=&\int d^{d+1} x \sqrt{-\hat G}\left\{a \hat R^2 
+ b \hat R_{\mu\nu}\hat R^{\mu\nu}
+ c \hat R_{\mu\nu\rho\sigma}\hat R^{\mu\nu\rho\sigma}
+ {1 \over \kappa^2} \hat R - \Lambda \right\} \nn
&& + S_B
\eea
Here $M_{d+1}$ is  $d+1$ dimensional manifold whose boundary is 
$d$ dimensional manifold $M_d$.
The boundary term $S_B$ is necessary to make the variational 
principle to be well-defined but we do not need the explicit expression 
of $S_B$ since it does not contribute to the anomaly.
Note that in Einstein gravity the boundary terms
 may be used to present the action as functional 
of fields and their first derivatives \cite{Gary}. One can
also add higher derivative boundary terms.  

Note that for some values of parameters above theory can be related 
with superstring or heterotic string effective action. Superstring 
effective action does not include $R^2$ terms in $d+1=10$ dimensions.
However, making Calabi-Yau compactification in lower dimensions 
one can get also $R^2$ terms (see for example \cite{AGN}). We may 
consider as particular example of higher derivative gravity the
effective action of heterotic string \cite{CFMP,Tse} which we take 
for torus compactification and supposing all fields except
the metric tensor to be constant.
\be
\label{sp1}
S_{het}=\int d^{d+1}x\sqrt{-g}\left\{{1 \over \kappa^2}R
-\Lambda +{\alpha' \over 8}
R_{\mu\nu\rho\sigma}R^{\mu\nu\rho\sigma}\right\}\ .
\ee
The cosmological constant may be non-zero, for example 
for covariantly constant gauge fields.

The conventions of curvatures are given by
\bea
\label{curv}
R&=&g^{\mu\nu}R_{\mu\nu} \nn
R_{\mu\nu}&=& -\Gamma^\lambda_{\mu\lambda,\kappa}
+ \Gamma^\lambda_{\mu\kappa,\lambda}
- \Gamma^\eta_{\mu\lambda}\Gamma^\lambda_{\kappa\eta}
+ \Gamma^\eta_{\mu\kappa}\Gamma^\lambda_{\lambda\eta} \nn
\Gamma^\eta_{\mu\lambda}&=&{1 \over 2}g^{\eta\nu}\left(
g_{\mu\nu,\lambda} + g_{\lambda\nu,\mu} - g_{\mu\lambda,\nu} 
\right)\ .
\eea

First we investigate if the equations of motion for the action 
(\ref{vi}) have a solution which describes anti de Sitter space, 
whose metric is given by
\be
\label{ai}
ds^2=\hat G^{(0)}_{\mu\nu}dx^\mu dx^\nu 
= {l^2 \over 4}\rho^{-2}d\rho d\rho + \sum_{i=1}^d
\rho^{-1} \eta_{ij}dx^i dx^j \ .
\ee
When we assume the metric in the form  (\ref{ai}), 
the scalar, Ricci and Riemann curvatures are given by
\be
\label{aii}
\hat R^{(0)}=-{d^2 + d \over l^2}\ ,\ \ 
\hat R^{(0)}_{\mu\nu}=-{d \over l^2}G^{(0)}_{\mu\nu}\ ,\ \ 
\hat R^{(0)}_{\mu\nu\rho\sigma}=-{1 \over l^2}
\left(G^{(0)}_{\mu\rho}G^{(0)}_{\nu\sigma}
-G^{(0)}_{\mu\sigma}G^{(0)}_{\nu\rho}\right)\ ,
\ee
which tell that these curvatures are covariantly constant. 
Then in the equations of motion from the action (\ref{vi}), 
the terms containing the covariant derivatives of the curvatures 
vanish and the equations have the following forms:
\bea
\label{aiii}
0&=&-{1 \over 2}G^{(0)}_{\zeta\xi}
\left\{a \hat R^{(0)2} 
+ b \hat R^{(0)}_{\mu\nu}\hat R^{(0)\mu\nu}
+ c \hat R^{(0)}_{\mu\nu\rho\sigma}\hat R^{(0)\mu\nu\rho\sigma}
+ {1 \over \kappa^2} \hat R^{(0)} - \Lambda \right\} \nn
&& + 2a R^{(0)} R^{(0)}_{\zeta\xi} 
+ 2b \hat R^{(0)}_{\mu\zeta}{\hat R^{(0)\mu}}_\xi
+ 2c \hat R^{(0)}_{\zeta\mu\nu\rho}\hat R_\xi^{(0)\mu\nu\rho}
+ {1 \over \kappa^2} \hat R^{(0)}_{\zeta\xi}\ .
\eea
Then  substituting Eqs.(\ref{aii}) into (\ref{aiii}), we find 
\bea
\label{aiv}
0&=&{a \over l^4}(d+1)d^2(d-3) + {b \over l^4}d^2(d-3) \nn
&& + {2c \over l^4}d(d-3) - {d(d-1) \over \kappa^2 l^2}
-\Lambda\ .
\eea
The equation (\ref{aiv}) can be solved with respect to $l^2$ 
if 
\be
\label{aiva}
{d^2(d-1)^2 \over \kappa^4}-4d(d-3)\left\{(d+1)da + db + 2c\right\}
\Lambda\geq 0
\ee
which can been found from the determinant in (\ref{aiv}).
Then we obtain
\be
\label{ll}
l^2=-{{d(d+1) \over \kappa^2}\pm \sqrt{
{d^2(d-1)^2 \over \kappa^4}-4d(d-3)\left\{(d+1)da + db + 2c\right\}
\Lambda} \over 2\Lambda}\ .
\ee
The sign in front of the root in the above equation 
may be chosen to be positive what
corresponds to the Einstein gravity ($a=b=c=0$).

The existence of the solution where the sign is negative 
tells that there can be anti de Sitter solution even if the 
cosmological constant $\Lambda$ is positive since $l^2$ can 
be positive if $(d-3)\left\{(d+1)da + db + 2c\right\}<0$. 
The solution describes the de Sitter or anti de Sitter space, 
depending on the overall sign of $l^2$. $l^2$ can be negative or 
positive 
depending from the choice of the parameters and sign in above 
equation. 

In order to calculate the conformal anomaly, we consider the 
fluctuations around the anti de Sitter space (\ref{ai}).
As in \cite{HS,NOads}, we assume the metric has the 
following form:
\bea
\label{i}
ds^2&\equiv&\hat G_{\mu\nu}dx^\mu dx^\nu 
= {l^2 \over 4}\rho^{-2}d\rho d\rho + \sum_{i=1}^d
\hat g_{ij}dx^i dx^j \nn
\hat g_{ij}&=&\rho^{-1}g_{ij}\ .
\eea
We should note that there is a redundancy in the expression of 
(\ref{i}). In fact, if we reparametrize the metric :
\be
\label{ia}
\delta\rho= \delta\sigma\rho\ ,\ \ 
\delta g_{ij}= \delta\sigma g_{ij}\ .
\ee
by a constant parameter $\delta\sigma$, the expression (\ref{i}) 
is invariant. The transformation (\ref{ia}) is nothing but the scale 
transformation on $M_d$.

In the parametrization in (\ref{i})
we find the expressions for the scalar curvature $\hat R$
\bea
\label{iii}
\hat R&=&-{d^2 + d \over l^2}+\rho R 
+{2(d-1) \rho \over l^2}g^{ij}g'_{ij}
+ {3\rho^2 \over l^2} g^{ij}g^{kl}g'_{ik}g'_{jl} \nn
&& - {4\rho^2 \over l^2}g^{ij}g''_{ij}
- {\rho^2 \over l^2}g^{ij}g^{kl}g'_{ij}g'_{kl} 
\eea
for Ricci tensor $\hat R_{\mu\nu}$
\bea
\label{iv}
\hat R_{\rho\rho}&=&-{d \over 4\rho^2} - {1 \over 2}g^{ij}g''_{ij} 
+ {1 \over 4}g^{ik}g^{lj}g'_{kl}g'_{ij} \nn
\hat R_{ij}&=&R_{ij} - {2\rho \over l^2}g''_{ij} 
+ {2\rho \over l^2}g^{kl}g'_{ki}g'_{lj} 
- {\rho \over l^2}g'_{ij}g^{kl}g'_{kl} \nn
&& - {2-d \over l^2}g'_{ij} + {1 \over l^2}g_{ij}g^{kl}g'_{kl} 
- {d \over l^2\rho}g_{ij} \nn
\hat R_{i\rho}&=& \hat R_{\rho i} \nn
&=&{1 \over 2}g^{jk}g'_{ki,j} - {1 \over 2}g^{kj}g'_{jk,i} 
+ {1 \over 2}g^{jk}_{,j}g'_{ki} \nn
&& + {1 \over 4}g^{kl}g'_{li}g^{jm}g_{jm,k} 
- {1 \over 4}g^{kj}_{,i}g'_{jk} 
\eea
and for the Riemann tensor $\hat R_{\mu\nu\rho\sigma}$
\bea
\label{v}
\hat R_{\rho\rho\rho\rho}&=& 0 \nn
\hat R_{\rho\rho\rho i}&=&R_{\rho\rho i\rho}=R_{\rho i\rho\rho}
=R_{i \rho\rho\rho}=0 \nn
\hat R_{ij\rho\rho}&=&\hat R_{\rho\rho ij}=0 \nn
\hat R_{i\rho j\rho}&=&\hat R_{\rho i\rho j}=
-\hat R_{i\rho \rho j}=\hat R_{\rho ij \rho} \nn
&=& - {1 \over 4\rho^3}g_{ij} + {1 \over 4\rho}g^{kl}g'_{ki}g'_{lj}
- {1 \over 2\rho}g''_{ij} \nn
\hat R_{\rho ijk}&=& - \hat R_{i \rho jk} = \hat R_{jk \rho i}
= - \hat R_{jki \rho} \nn
&=& {1 \over 4\rho }\left\{ 2g'_{ij,k} - 2g'_{ik,j} 
- g^{lm}\left(g_{im,k} + g_{km,i} - g_{ik,m}\right)g'_{lj} 
\right. \nn
&& \left. + g^{lm}\left(g_{im,j} + g_{jm,i} - g_{ij,m}\right)g'_{lk} 
\right\} \nn
\hat R_{ijkl}&=&{1 \over \rho}R_{ijkl} \\
&& - {1 \over \rho^2 l^2}\left\{\left( g_{jl}-\rho g'_{jl}\right)
\left(g_{ik} - \rho g'_{ik}\right)
-\left( g_{jk}-\rho g'_{jk}\right)
\left(g_{il} - \rho g'_{il}\right)\right\} \nonumber
\eea
Here ``$\ '\ $" expresses the derivative with respect to $\rho$ 
and $R$, $R_{ij}$, $R_{ijkl}$ are scalar curvature, Ricci and 
Riemann tensors, respectively, on $M_d$.

As in the previous papers \cite{HS,NOads}, 
we expand the metric $g_{ij}$ as a power 
series with respect to $\rho$,
\be
\label{vii}
g_{ij}=g_{(0)ij}+\rho g_{(1)ij}+\rho^2 g_{(2)ij}+\cdots \ .
\ee
Substituting (\ref{vii}) into 
(\ref{iii}), (\ref{iv}), 
and (\ref{v}), we find the following expansions with respect to 
$\rho$:
\bea
\label{bi}
\sqrt{-\hat G}&=&{l \over 2}\rho^{-{d \over 2}-1}
\sqrt{-g_{(0)}}\left\{1 + {\rho \over 2}g_{(0)}^{ij}g_{(1)ij} 
\right. \nn
&& + \rho^2\left({1 \over 2}g_{(0)}^{ij}g_{(2)ij}
- {1 \over 4}g_{(0)}^{ij}g_{(0)}^{kl}g_{(1)ik} g_{(1)jl} \right.\nn
&& \left.\left. 
+ {1 \over 8}\left(g_{(0)}^{ij}g_{(1)ij} \right)^2\right)
+ O(\rho^3)\right\} \nn
\sqrt{-\hat G}\hat R&=&{l \over 2}\rho^{-{d \over 2}-1}
\sqrt{-g_{(0)}}\left\{-{d^2 + d \over l^2} \right. \nn
&& + \rho \left(R_{(0)}
+ {-d^2+3d-4 \over 2l^2}g_{(0)}^{ij}g_{(1)ij}\right)  \nn
&& + \rho^2\left( -g_{(1)ij}R_{(0)}^{ij}
+ {1 \over 2}R_{(0)}g_{(0)}^{ij}g_{(1)ij} \right. \nn
&& + {-d^2 + 7d -24 \over 2l^2}g_{(0)}^{ij}g_{(2)ij} \nn
&& + {d^2 - 7d + 20 \over 4l^2}
g_{(0)}^{ij}g_{(0)}^{kl}g_{(1)ik} g_{(1)jl} \nn
&& \left.\left. 
+ {-d^2 + 7d - 16 \over 8l^2}\left(g_{(0)}^{ij}g_{(1)ij} \right)^2
\right)+ O(\rho^3)\right\} \nn
\sqrt{-\hat G}\hat R^2&=&{l \over 2}\rho^{-{d \over 2}-1}
\sqrt{-g_{(0)}}\left\{{d^2(d+1)^2 \over l^2} \right. \nn
&& + \rho \left(- {2d(d+1) \over l^2}R_{(0)}
+ {d^4 -6d^3 + d^2 + 8d \over 2l^4}g_{(0)}^{ij}g_{(1)ij}\right) \nn
&& + \rho^2\left( R_{(0)}^2 +
{2d(d+1) \over l^2}g_{(1)ij}R_{(0)}^{ij} \right. \nn
&& + {-d^2 + 3d -4 \over l^2}R_{(0)}g_{(0)}^{ij}g_{(1)ij} \nn
&& + {d^4 -14d^3 +33d^2 + 48d \over 2l^4}g_{(0)}^{ij}g_{(2)ij} \nn
&& + {-d^4 + 14d^3 -25d^2 - 40d \over 4l^4}
g_{(0)}^{ij}g_{(0)}^{kl}g_{(1)ik} g_{(1)jl} \nn
&& \left.\left. + {d^4 -14d^3 +49d^2 -32d + 32 \over 8l^4}
\left(g_{(0)}^{ij}g_{(1)ij} \right)^2
\right)+ O(\rho^3)\right\} \nn
\sqrt{-\hat G}\hat R_{\mu\nu} \hat R^{\mu\nu}
&=&{l \over 2}\rho^{-{d \over 2}-1}
\sqrt{-g_{(0)}}\left\{{d^2(d+1) \over l^2} \right. \nn
&& + \rho \left(- {2d \over l^2}R_{(0)}
+ {d^3 -7d^2 + 8d \over 2l^4}g_{(0)}^{ij}g_{(1)ij}\right) \nn
&& + \rho^2\left( R_{(0)}^{ij}R_{(0)ij} 
+{4d -4 \over l^2}g_{(1)ij}R_{(0)}^{ij} \right. \nn
&& + {-d +2 \over l^2}R_{(0)}g_{(0)}^{ij}g_{(1)ij} 
+ {d^3 -15d^2 + 48d  \over 2l^4}g_{(0)}^{ij}g_{(2)ij} \nn
&& + {-d^3 +19d^2 - 56d +16 \over 4l^4}
g_{(0)}^{ij}g_{(0)}^{kl}g_{(1)ik} g_{(1)jl} \nn
&&  \left.\left. + {d^3 -15d^2 +56d -32 \over 8l^4}
\left(g_{(0)}^{ij}g_{(1)ij} \right)^2
\right)+ O(\rho^3)\right\} \nn
\sqrt{-\hat G}\hat R_{\mu\nu\rho\sigma} \hat R^{\mu\nu\rho\sigma}
&=&{l \over 2}\rho^{-{d \over 2}-1}
\sqrt{-g_{(0)}}\left\{{2d(d+1) \over l^4} \right. \nn
&& + \rho \left(- {4 \over l^2}R_{(0)}
+ {d^2 -7d +8 \over 2l^4}g_{(0)}^{ij}g_{(1)ij}\right)  \nn
&& + \rho^2\left( R_{(0)}^{ijkl}R_{(0)ijkl} 
+ {4 \over l^2}g_{(1)ij}R_{(0)}^{ij}
- {2 \over l^2}R_{(0)}g_{(0)}^{ij}g_{(1)ij} \right. \nn
&& + {d^2 -15d +48 \over l^4}g_{(0)}^{ij}g_{(2)ij} \nn
&& + {-d^2 +23d -56 \over 2l^4}
g_{(0)}^{ij}g_{(0)}^{kl}g_{(1)ik} g_{(1)jl} \nn
&& \left.\left. + {d^2 -15d + 48 \over 4l^4}
\left(g_{(0)}^{ij}g_{(1)ij} \right)^2
\right) + O(\rho^3)\right\} \ .
\eea

We regard $g_{(0)ij}$ in (\ref{vii}) as independent field on 
$M_d$. We can solve $g_{(l)ij}$ ($l=1,2,\cdots$) with respect to 
$g_{(0)ij}$  using equations of motion.
When substituting the expression (\ref{vii}) or (\ref{bi}) 
into the classical action (\ref{vi}), 
the action diverges in general since the action contains the 
infinite volume integration on $M_{d+1}$. 
We regularize the infrared divergence by introducing a 
cutoff parameter $\epsilon$(following to Witten \cite{W}):
\be
\label{viii}
\int d^{d+1}x\rightarrow \int d^dx\int_\epsilon d\rho \ ,\ \ 
\int_{M_d} d^d x\Bigl(\cdots\Bigr)\rightarrow 
\int d^d x\left.\Bigl(\cdots\Bigr)\right|_{\rho=\epsilon}\ .
\ee
Then the action (\ref{vi}) can be expanded as a power series of 
$\epsilon$:
\bea
\label{viiia}
S&=&S_0(g_{(0)ij})\epsilon^{-{d \over 2}}
+ S_1(g_{(0)ij}, g_{(1)ij})\epsilon^{-{d \over 2}-1} \nn
&& + \cdots + S_{\rm ln} \ln \epsilon 
+ S_{d \over 2} + {\cal O}(\epsilon^{1 \over 2}) \ .
\eea
The term $S_{\rm ln}$ proportional to $\ln\epsilon$ 
appears when $d=$even. 
In (\ref{viiia}), the terms proportional to the inverse power of 
$\epsilon$ in the regularized action are invariant under the scale 
transformation 
\be
\label{viiib}
\delta g_{(0)\mu\nu}=2\delta\sigma g_{(0)\mu\nu}\ ,\ \  
\delta\epsilon=2\delta\sigma\epsilon \ . 
\ee
The invariance comes from the redanduncy (\ref{ia}). 
The subtraction of these terms proportional to the 
inverse power of $\epsilon$ does not break the invariance.
When $d$ is even, however, there appears the term 
$S_{\rm ln}$ proportional to $\ln\epsilon$.
The subtraction of the term $S_{\rm ln}$ breaks 
the invariance under the transformation(\ref{viiib}). 
The reason is that the variation of the $\ln\epsilon$ term 
under the scale transformation (\ref{viiib}) is finite 
when $\epsilon\rightarrow 0$ since 
$\ln\epsilon \rightarrow \ln\epsilon + \ln(2\delta\sigma)$. 
Therefore the variation should be canceled by the variation of 
the finite term $S_{d \over 2}$ (which does not depend on $\epsilon$) 
\be
\label{vari}
\delta S_{d \over 2}=-\ln (2\sigma)S_{\rm ln}
\ee
since the original total action (\ref{vi}) is invariant under the 
scale transformation. 
Since the action $S_{d \over 2}$ can be regarded as the action 
renormalized by the subtraction of the terms which diverge when 
$\epsilon\rightarrow 0$, 
the $\ln\epsilon$ term $S_{\rm ln}$ gives the conformal anomaly $T$ 
of the renormalized theory on the boundary $M_d$: 
\be
\label{xi}
S_{\rm ln}=-{1 \over 2}
\int d^2x \sqrt{-g_{(0)}}T \ .
\ee

First we consider $d=2$ case.
When $d=2$,  substituting (\ref{vii}) or (\ref{bi}) into 
the action (\ref{vi}) and using the regularization (\ref{viii}), 
we obtain
\bea
\label{ix}
S_{\rm ln}&=&
\int d^2 x \sqrt{-g_{(0)}}\left[
g_{(0)}^{ij}g_{(1)ij}\left\{ -{3a \over l^3}\right.\right. \nn
&& \left. - {b \over l^3} - {c \over l^3} 
- {1 \over 2l\kappa^2} - {l\Lambda \over 4} \right\} \nn
&& \left. + R_{(0)}\left\{ -{6 a \over l}-{2b \over l} - {2c \over l}
+ {1 \over 2\kappa^2}\right\}\right]\ .
\eea
Since Eq.(\ref{aiv}) gives the following equation for $d=2$
\be
\label{aiv2}
0=-{12a \over l^4} - {4b \over l^4} - {4c \over l^4}
- {2 \over \kappa^2 l^2} - \Lambda \ ,
\ee
we find the term proportional to $g_{(0)}^{ij}g_{(1)ij}$ in (\ref{ix}) 
vanishes. Then  using (\ref{xi}), we find the expression 
of Weyl anomaly $T$
\be
\label{xii}
T= \left\{ 12a + 4b + 4c 
- {l \over \kappa^2}\right\}R_{(0)}\ .
\ee
When $a=b=c=0$, the above equation reproduces the result 
of refs.\cite{HS,NOads}:
\be
\label{xiiHS}
T= - {l \over \kappa^2}R_{(0)}\ .
\ee
For the heterotic string effective action (\ref{sp1}), 
we obtain
\be
\label{xiiST}
T_{\rm string}= \left\{ {\alpha' \over 2}
- {l \over \kappa^2}\right\}R_{(0)}\ .
\ee

The important result which follows from above consideration 
is the following:
two-dimensional CFT may have description not only in terms 
of three-dimensional Einstein gravity (bulk theory) but also
in terms of three-dimensional higher derivative gravity  where 
Einstein term may be absent (but cosmological term is present).
Choosing the coefficients $a,b,c$ in Eq.(\ref{xii}) so that 
the correspondent conformal anomaly would give the 
anomaly for two-dimensional scalar or spinor CFT we get 
new examples of AdS/CFT correspondence. It is interesting that
in this case new theory appears on the bulk side. Note that
one can consider higher derivative gravity with Einstein term but
without of cosmological term as the bulk theory. 

\section{Conformal anomaly for $d=4$ theory}

In this section, we consider $d=4$ case.
 Substituting the expressions in (\ref{bi}) into the 
action (\ref{vi}) and using the regularization in (\ref{viii}), 
we find 
\bea
\label{xiii}
S_{\rm ln}&=&-{1 \over 2}\int dx^4\sqrt{-g_{(0)}}\left[
l \left(aR_{(0)}^2 + bR_{(0)}^{ij}R_{(0)ij}
+ c R_{(0)}^{ijkl}R_{(0)ijkl}\right) \right. \nn
&& + \left( {40a \over l^3} + {8b \over l^3} 
+ {4c \over l^3} - {6 \over l \kappa^2} - {l\Lambda \over 2}\right)
g_{(0)}^{ij}g_{(2)ij} \nn
&& + \left({40a \over l} + {12b \over l} + {12c \over l} 
- {l \over \kappa^2}\right)g_{(1)ij}R_{(0)}^{ij} \nn
&& + \left(-{8a \over l} - {2b \over l} - {2c \over l} 
+ {l \over 2\kappa^2}\right)R_{(0)}g_{(0)}^{ij}g_{(1)ij} \nn
&& + \left({20a \over l^3} + {8b \over l^3} 
+ {10c \over l^3} + {2 \over l\kappa^2} + {l\Lambda \over 4}\right)
g_{(0)}^{ij}g_{(0)}^{kl}g_{(1)ik}g_{(1)jl} \nn
&& \left. + \left({6a \over l^3} + {2b \over l^3} 
+ {c \over l^3} - {1 \over 2l\kappa^2} - {l\Lambda \over 8}\right)
\left(g_{(0)}^{ij}g_{(1)ij}\right)^2 \right]
\eea
Other terms proportional $\epsilon^{-1}$ or $\epsilon^{-2}$ 
which diverge when $\epsilon\rightarrow 0$ can be subtracted without 
loss of the general covariance and scale invariance.
Since Eq.(\ref{aiv}) has the following form in $d=4$,
\be
\label{aiv4}
0={80a \over l^4} + {16b \over l^4} + {8c \over l^4}
- {12 \over \kappa^2 l^2} - \Lambda \ ,
\ee
we find that 
the term in (\ref{xiii}) proportional to $g_{(0)}^{ij}g_{(2)ij}$ 
vanishes.
The equation given by the variation over $g_{(1)ij}$ is given by
\bea
\label{g1i}
0&=&AR_{(0)}^{ij} + B g_{(0)}^{ij}R_{(0)} 
+ 2C g_{(0)}^{ik}g_{(0)}^{jl}g_{(1)kl}
+ 2D g_{(0)}^{ij}g_{(0)}^{kl}g_{(1)kl} \nn
A&\equiv&{40a \over l} + {12b \over l} + {4c \over l} 
- {l \over \kappa^2} \nn
B&\equiv&-{8a \over l} - {2b \over l} - {2c \over l} 
+ {l \over 2\kappa^2} \nn
C&\equiv& {20a \over l^3} + {8b \over l^3} 
+ {10c \over l^3} + {2 \over l\kappa^2} + {l\Lambda \over 4} 
={40a \over l^3} + {12b \over l^3} 
+ {12c \over l^3} - {1 \over l\kappa^2}
\nn
D&\equiv& {6a \over l^3} + {2b \over l^3} 
+ {c \over l^3} - {1 \over 2l\kappa^2} - {l\Lambda \over 8}
=-{4a \over l^3}  + {1 \over l\kappa^2} \ .
\eea
Here we used (\ref{aiv4}) to re-write the expressions of $C$ and $D$ 
and to remove the cosmological constant $\Lambda$. Multiplying 
$g_{(1)ij}$ with (\ref{g1i}), we obtain
\be
\label{g1ii}
g_{(0)}^{ij} g_{(1)ij}=-{A+4B \over 2(C+4D)}R_{(0)}\ .
\ee
Substituting (\ref{g1ii}) into (\ref{g1i}), we 
can solve (\ref{g1i}) with respect to $g_{(1)ij}$ as follows:
\be
\label{xiv}
g_{(1)ij}=-{A \over 2C}R_{(0)ij}
+ {AD - BC \over 2C(C+4D)}R_{(0)}g_{(0)ij} \ .
\ee
Substituting (\ref{xiv}) into (\ref{xiii}) (and  using 
(\ref{aiv4})), we find the following expression for the anomaly $T$
\bea
\label{xv}
T&=&\left( al + 
{A^2D - 4B^2C - 2ABC \over 4C(C+4D) }\right) R_{(0)}^2 \nn
&& + \left(bl - {A^2 \over 4C}\right)
R_{(0)ij}R_{(0)}^{ij} + cl R_{(0)ijkl}R_{(0)}^{ijkl}\ .
\eea
If we put $a=b=c=0$, the above expression reproduces the result in 
\cite{HS,NOads}:
\be
\label{xvHS}
T=-{l^3 \over 12\kappa^2} R_{(0)}^2 
+ {l^3 \over 4\kappa^2}R_{(0)ij}R_{(0)}^{ij} \ .
\ee
For general case, by substituting (\ref{g1i}) into (\ref{xv}), 
we obtain
\be
\label{xvaa}
T=\left(-{l^3 \over 8\kappa^2}+5al+bl\right)(G-F) 
+ {cl \over 2}(G+F)\ .
\ee
Here we used the Gauss-Bonnet 
invariant $G$ and the square of the Weyl tensor $F$, 
which are given by
\bea
\label{GF} 
G&=&R_{(0)}^2 -4 R_{(0)ij}R_{(0)}^{ij} 
+ R_{(0)ijkl}R_{(0)}^{ijkl} \nn
F&=&{1 \over 3}R_{(0)}^2 -2 R_{(0)ij}R_{(0)}^{ij} 
+ R_{(0)ijkl}R_{(0)}^{ijkl} \ .
\eea

This is really beautiful and compact result which shows 
that apparently any variant of higher derivative AdS gravity 
has correspondence with some $d=4$ conformal field theory. 
It is very general and it maybe used in various contexts as we 
see in next section. It maybe also used to search new examples
of bulk/boundary correspondence with various higher derivative 
supergravities on bulk side. 

 Consider one particular example:
\be
\label{s6}
c=0\ ,\quad 
{l^3 \over \kappa^2}- 40al - 8bl ={2N^2 \over (4\pi)^2}\ ,
\ee 
Then,
the anomaly of $D=4$ ${\cal N}=4$ super Yang-Mills 
is reproduced. 

It is very interesting that last relation is still true 
even in case of absence the Einstein term there. In fact, 
in this case we found new example of AdS/CFT correspondence
where boundary QFT is well-known ${\cal N}=4$ super Yang-Mills 
theory and bulk side is described by higher derivative 
five-dimensional gravity with cosmological term and without
(or with) the Einstein term. Moreover, one can consider 
more narrow class of theories with $a=0$ or $b=0$ in 
initial classical action.
 
It should be also noted that one cannot consider pure 
higher derivative gravity in above context because without 
cosmological and Einstein terms the theory is scale invariant.
Then it presumbly cannot be useful as bulk side in AdS/CFT 
correspondence.
In our approach the manifestation of this fact is that for 
pure $R^2$-gravity the conformal anomaly is zero.

Finally we note that when only $c$ and cosmological constant 
are not zero among the parameters of AdS higher derivative 
gravity as it follows from general Eq.(\ref{xvaa}) bulk 
side does not correspond to ${\cal N}=4$ super Yang-Mills theory. 

\section{Next-to-leading corrections to ${\cal N}=2$ 
superconformal theory trace anomaly from AdS/CFT 
correspondence}

Our main motivativation in this work was to find new higher 
derivative versions of bulk theory in AdS/CFT correspondence 
via calculation of holografic conformal anomaly. However, 
our general result maybe applied perfectly well for
 another formulations (or motivation): when higher 
derivatives terms in AdS gravity appear as next-to-
leading corrections in low-energy string effective action 
(after compactification). Let us demonstrate this on 
the example of ${\cal N}=2$ SCFT (and corresponding bulk
AdS higher derivative gravity).

Recently (after the first version of this paper appeared 
in hep-th ),  
the trace anomaly of $d=4$, ${\cal N}=2$ and 
${\cal N}=4$ SCFT has been 
investigated up to next to leading order in the $1/N$ 
expansion by Blau, Narain and Gava \cite{BNG}. 
For an ${\cal N}=2$ SCFT theory with $n_V$ vector multiplets 
and $n_H$ hypermultiplets, the trace anomaly is, by usual 
UV calculations, given by
\bea
\label{bng1}
T&=&{1 \over 24\cdot 16\pi^2}\Bigl[ -{1 \over 3}
(11n_V + n_H) R^2 \nn
&& + 12n_V R_{ij}R^{ij} + (n_H-n_V)R_{ijkl}R^{ijkl}\Bigr]\ ,
\eea
and especially, when the gauge group is $Sp(N)$, 
\bea
\label{bng2}
T&=&{1 \over 24\cdot 16\pi^2}\Bigl[ 
-\left(8N^2 + 6N - {1 \over 3}\right) R^2 \nn
&& + (24N^2+12N) R_{ij}R^{ij} + (6N-1)R_{ijkl}R^{ijkl}\Bigr]\ .
\eea
The ${\cal N}=2$ theory with the gauge group $Sp(N)$ arise as 
the low-energy theory on the world volume on $N$ D3-branes 
sitting inside 8 D7-branes at an O7-plane \cite{Sen}. 
The string theory dual to this theory has been conjectured 
to be type IIB string theory on $AdS_5\times X^5$ where 
$X_5=S^5/Z_2$ \cite{FS}, whose low energy effective action 
is given by (see the corresponding derivation in ref.\cite{BNG}) 
\be
\label{bng3} 
S=\int_{AdS_5} d^5x \sqrt{G}\left\{{N^2 \over 4\pi^2}
\left(R-2\Lambda\right) 
+ {6N \over 24\cdot 16\pi^2}R_{\mu\nu\rho\sigma}
R^{\mu\nu\rho\sigma}\right\}\ .
\ee
In order to use the general result for the conformal anomaly 
$T$ in (\ref{xvaa}), we compare (\ref{bng3}) with (\ref{iv}) 
and find 
\be
\label{bng4}
{1 \over \kappa^2}={N^2 \over 4\pi^2}\ , \quad
a=b=0\ , \quad c={6N \over 24\cdot 16\pi^2}\ .
\ee
Futhermore we need to choose
\be
\label{bng5}
\Lambda=-{12 \over \kappa^2}=-{12N^2 \over 4\pi^2}\ ,
\ee
which corresponds to that the length scale is chosen to be 
unity in \cite{BNG}. We should note, however, the length scale 
$l$ in our paper has a correction.  Substituting (\ref{bng4}) 
and (\ref{bng5}) into (\ref{aiv4}), we find 
\bea
\label{bng6}
{1 \over l^2}&=&1+{2c\kappa^2 \over 3l^2} \nn
&=&1+{2c\kappa^2 \over 3} + {\cal O}\left((c\kappa^2)^2\right)\ .
\eea
In the second line in (\ref{bng6}), we have assumed 
$c\kappa^2={1 \over 16N}$ is small or $N$ is large.
Then we find the following expression for $T$:
\bea
\label{bng7}
T&=&\left(-{1 \over 8\kappa^2}+{c \over 8}\right)(G-F) 
+ {c \over 2}(G+F) + {1 \over \kappa^2}{\cal O}
\left((c\kappa)^2\right) \nn
&=&{N^2 \over 16\pi^2}\left(-{1 \over 3}R_{(0)}^2 
+ R_{(0)ij}R_{(0)}^{ij}\right) \nn
&& + {6N \over 24\cdot 16\pi^2}\left({3 \over 4}R_{(0)}^2 
-{13 \over 4}R_{(0)ij}R_{(0)}^{ij} + R_{(0)ijkl}R_{(0)}^{ijkl}
\right) + {\cal O}(1)\ .
\eea
In the first line in (\ref{bng7}), 
we have substituted (\ref{bng6}) and put $a=b=0$ and 
in the second line, (\ref{bng4}), (\ref{bng5}) and the explicit 
expression for $G$ and $F$ in (\ref{GF}) are substituted.
Eq. (\ref{bng7}) exactly reproduces the result in (\ref{bng1}) 
as in \cite{BNG}. 

Hence, we demonstrated how to apply our general 
formalism to calculate next-to-leading order corrections 
to conformal anomaly from bulk AdS side. Similarly, one can 
get conformal anomaly from bulk side for any other 
low-energy string effective action.

\section{Discussion}

In summary, we studied higher derivative gravity as bulk
theory in AdS/CFT correspondence. We explicitly calculated 
the conformal anomaly from such theory following to Witten 
proposal and assuming that they describe some boundary QFTs.
This approach opens new lines in AdS/CFT correspondence as
it presents new versions of bulk theory.
For example, three-dimensional $R^2$ gravity with cosmological 
constant and with (or without) Einstein term may describe 
$d=2$ scalar or spinor CFT. In the same way,
five-dimensional $R^2$ gravity with cosmological constant 
and with (or without) Einstein term may be considered as 
bulk theory for ${\cal N}=4$ super Yang-Mills quantum theory.
Pure  $R^2$ gravity seems to be not good as bulk theory 
due to its scale invariance (no mass scale). We also applied 
our general result to calculation of next-to-leading 
corrections to conformal anomaly of ${\cal N}=4$ SCFT from 
bulk AdS higher derivative gravity. Perfect agreement with results 
of very recent work \cite{BNG} is found.

There are also interesting possibilities to extend 
the results of present work. In the recent paper \cite{GRW} 
it was studied the conformal anomaly of submanifold observables 
in AdS/CFT correspondence. Such investigation is expected to be useful 
for the study of Wilson loops in large $N$ theories in frames 
of bulk-boundary correspondence\cite{MRY}. It is clear that modification 
of bulk theory (say of the form we discuss in this work) 
will also modify the conformal anomaly on submanifolds. 

In the recent paper\cite{BK} the form of Witten proposal has 
been applied to study the boundary stress tensor of anti-de Sitter 
Einstein gravity in $d+1=3$ and $d+1=5$ dimensions via 
AdS/CFT correspondence. As by-product the conformal anomaly
in $d=2$ and $d=4$ is also recovered. It would be of interest 
to re-consider this problem for our version of higher derivative 
gravity as it may lead to acceptable definition of the gravitational 
energy and stress tensor even in such case.

For example, the Casimir energy of $d=4$ $SU(N)$ ${\cal N}=4$ 
super Yang-Mills theory on $S^3\times R$ 
which is global boundary of five-dimensional AdS is given by 
\be
\label{d1}
E_{\rm Casimir}={3(N^2 - 1) \over 16 r}\ .
\ee
Here $r$ is the radius of $S^3$. By multiplying 
$\sqrt{-g_{tt}}=r/l$, the Casimir mass is given by
\be
\label{d2}
M_{\rm Casimir}={3(N^2 - 1) \over 16 l}\ ,
\ee
which gives the non-vanishing ground state energy for AdS${_5}$,
i.e. corresponding BH mass \cite{BK}. 
Using eq.(\ref{s6}) 
as the conversion formula to gauge theory variables 
from higher derivative gravity we get AdS mass for higher 
derivative gravity suggesting the way to construct stress-energy 
tensor in such theory.

As the final remark we would like to point out that Eq.(\ref{xvaa}) 
maybe used in the constructive way in order to find new examples 
of bulk/boundary relations. Let us imagine that we got some variant 
of low-energy string effective action (after some compactification) 
as particular version of $d=5$ AdS higher derivative gravity.
It may also come in another way as bosonic sector of some 
higher deivative AdS supergravity. Eq.(\ref{xvaa}) gives 
holografic conformal anomaly (up to next-to-leading terms) 
from bulk side. Having such conformal anomaly maybe often enough to
suggest new boundary CFT.

\ 

\noindent
{\bf Acknowledgments}. We thank M. Blau, K. Narain and 
E. Gava for pointing out mistake in the first version of this work.

\end{document}